\documentclass[%
superscriptaddress,
groupedaddress,
preprint,
nofootinbib,
amsmath,amssymb,
aps,
longbibliography,
]{revtex4-2}

\usepackage{graphicx}% Include figure files
\usepackage{dcolumn}% Align table columns on decimal point
\usepackage{bm}% bold math
\usepackage{xcolor}

\usepackage{hyperref}% add hypertext capabilities
\usepackage[mathlines]{lineno}% Enable numbering of text and display math
\newcommand{\uB}{MicroBooNE}
\newcommand{\mB}{MiniBooNE}

\definecolor{cmntblue}{rgb}{0.0, 0.58, 0.71}
\definecolor{cmntgreen}{rgb}{0.0, 0.42, 0.24}
\definecolor{cmntbrown}{rgb}{0.6, 0.3, 0.2}

\usepackage[final, inline, nomargin]{fixme}
\fxsetup{theme=color, mode=multiuser}
\FXRegisterAuthor{bl}{1}{\color{blue}BL}  
\FXRegisterAuthor{tl}{2}{\color{red}TL}
\FXRegisterAuthor{nb}{3}{\color{cmntblue}NB}
\FXRegisterAuthor{pm}{4}{\color{cmntgreen}PM}
\FXRegisterAuthor{kh}{5}{\color{green}KH}
\FXRegisterAuthor{pts}{6}{\color{cmntbrown}PTS}
\FXRegisterAuthor{btf}{7}{\color{teal}BTF}
\newcommand{\nuebar}{\ensuremath{\overline{\nu}_{e}}}
\newcommand{\uFive}{$^{235}$U}

\newcommand{\pNine}{$^{239}$Pu}

\newcommand{\dmTwo}{$\Delta m^2$}

\newcommand{\NIST}{National Institute of Standards and Technology (NIST) \renewcommand{\NIST}{NIST}}
\newcommand{\NBSR}{National Bureau of Standards Reactor (NBSR) \renewcommand{\NBSR}{NBSR}}
\newcommand{\INL}{Idaho National Laboratory (INL) \renewcommand{\INL}{INL}}
\newcommand{\ATR}{Advanced Test Reactor (ATR) \renewcommand{\ATR}{ATR}}
\newcommand{\ORNL}{Oak Ridge National Laboratory (ORNL) \renewcommand{\ORNL}{ORNL}}
\newcommand{\HFIR}{High Flux Isotope Reactor (HFIR) \renewcommand{\HFIR}{HFIR}}
\newcommand{\LLNL}{Lawrence Livermore National Laboratory (LLNL) \renewcommand{\LLNL}{LLNL}}

\newcommand\snowmass{\begin{center}\rule[-0.2in]{\hsize}{0.01in}\\\rule{\hsize}{0.01in}\\
\vskip 0.1in Submitted to the  Proceedings of the US Community Study\\ 
on the Future of Particle Physics (Snowmass 2021)\\ 
\rule{\hsize}{0.01in}\\\rule[+0.2in]{\hsize}{0.01in} \end{center}}

\begin{document}

% \preprint{APS/123-QED}

\title{Physics Opportunities with PROSPECT-II}% Force line breaks with \\
 % !TEX root = main.tex

\author{M~Andriamirado$^{6}$,
A~B~Balantekin$^{17}$,
C~D~Bass$^{8}$,
D~E~Bergeron$^{9}$,
E~P~Bernard$^{7}$,
N~S~Bowden$^{7}$,
C~D~Bryan$^{10}$,
R~Carr$^{ 15}$, 
T~Classen$^{7}$,
A~J~Conant$^{10}$,
G~Deichert$^{10}$,
A~Delgado$^{11,14}$,
M~V~Diwan$^{2}$,
M~J~Dolinski$^{3}$,
A~Erickson$^{4}$,
B~T~Foust$^{18}$,
J~K~Gaison$^{18}$,
A~Galindo-Uribari$^{11,14}$,
C~E~Gilbert$^{11}$,
S~Gokhale$^{2}$,
C~Grant$^{1}$, 
S~Hans$^{2}$,
A~B~Hansell$^{12}$,
K~M~Heeger$^{18}$,
B~Heffron$^{11,14}$,
D~E~Jaffe$^{2}$,
S~Jayakumar$^{3}$,
X~Ji$^{2}$,
D~C~Jones$^{13}$, 
J~Koblanski$^{5}$,
P~Kunkle$^{1}$, 
O~Kyzylova$^{3}$, 
C~E~Lane$^{3}$,
T~J~Langford$^{18}$,
J~LaRosa$^{9}$, 
B~R~Littlejohn$^{6}$,
X~Lu$^{11,14}$, 
J~Maricic$^{5}$,
M~P~Mendenhall$^{7}$,
A~M~Meyer$^{5}$, 
R~Milincic$^{5}$, 
P~E~Mueller$^{11}$,
H~P~Mumm$^{9}$,
J~Napolitano$^{13}$,
R~Neilson$^{3}$,
J~A~Nikkel$^{18}$, 
S~Nour$^{9}$, 
J~L~Palomino$^{6}$, 
D~A~Pushin$^{16}$,
X~Qian$^{2}$,
C~Roca$^{7}$,
R~Rosero$^{2}$,
M~Searles$^{10}$,
P~T~Surukuchi$^{18}$,
F~Sutanto$^{7}$,
M~A~Tyra$^{9}$, 
R~L~Varner$^{11}$,
D~Venegas-Vargas$^{11,14}$, 
P~B~Weatherly$^{3}$, 
J~Wilhelmi$^{18}$,
A~Woolverton$^{16}$, 
M~Yeh$^{2}$,
C~Zhang$^{2}$ and
X~Zhang$^{7}$ \\
% (The PROSPECT Collaboration)
}

\address{$^{1}$Department of Physics, Boston University, Boston, MA, USA} \vspace{-0.4\baselineskip}
\address{$^{2}$Brookhaven National Laboratory, Upton, NY, USA} \vspace{-0.4\baselineskip}
\address{$^{3}$Department of Physics, Drexel University, Philadelphia, PA, USA} \vspace{-0.4\baselineskip}
\address{$^{4}$George W.\,Woodruff School of Mechanical Engineering, Georgia Institute of Technology, Atlanta, GA, USA} \vspace{-0.4\baselineskip}
\address{$^{5}$Department of Physics and Astronomy, University of Hawaii, Honolulu, HI, USA} \vspace{-0.4\baselineskip}
\address{$^{6}$Department of Physics, Illinois Institute of Technology, Chicago, IL, US} \vspace{-0.4\baselineskip}
\address{$^{7}$Nuclear and Chemical Sciences Division, Lawrence Livermore National Laboratory, Livermore, CA, USA} \vspace{-0.4\baselineskip}
\address{$^{8}$Department of Physics, Le Moyne College, Syracuse, NY, USA} \vspace{-0.4\baselineskip}
\address{$^{9}$National Institute of Standards and Technology, Gaithersburg, MD, USA} \vspace{-0.4\baselineskip}
\address{$^{10}$High Flux Isotope Reactor, Oak Ridge National Laboratory, Oak Ridge, TN, USA} \vspace{-0.4\baselineskip}
\address{$^{11}$Physics Division, Oak Ridge National Laboratory, Oak Ridge, TN, USA} \vspace{-0.4\baselineskip}
\address{$^{12}$Department of Physics, Susquehanna University, Selinsgrove, PA, USA} \vspace{-0.4\baselineskip}
\address{$^{13}$Department of Physics, Temple University, Philadelphia, PA, USA} \vspace{-0.4\baselineskip}
\address{$^{14}$Department of Physics and Astronomy, University of Tennessee, Knoxville, TN, USA} \vspace{-0.4\baselineskip}
\address{$^{15}$Department of Physics, United States Naval Academy, Annapolis, MD, USA} \vspace{-0.4\baselineskip}
\address{$^{16}$Institute for Quantum Computing and Department of Physics, University of Waterloo, Waterloo, ON, Canada} \vspace{-0.4\baselineskip}
\address{$^{17}$Department of Physics, University of Wisconsin, Madison, WI, USA} \vspace{-0.4\baselineskip}
\address{$^{18}$Wright Laboratory, Department of Physics, Yale University, New Haven, CT, USA} \vspace{-0.4\baselineskip}

%\ead{prospect.collaboration@gmail.com}
\collaboration{PROSPECT Collaboration}\email{prospect.collaboration@gmail.com}
\date{\today}% It is always \today, today,

% \begin{abstract}
% With the release of new results from \uB{} and BEST, the search for sterile neutrinos has grown even more complicated and interesting. 
% While \uB{} first results do not show an excess of electron-like signatures similar to that observed by \mB{}, their current sensitivity does not exclude all of sterile neutrino oscillation parameter space suggested by the \mB{} result.  
% In fact, in the electron neutrino channel, \uB{} potentially shows indications of an oscillation-based deficit. 
% Alternatively, the BEST experiment's recent results confirm the Gallium Anomaly and further increases its significance.
% No strong particle or nuclear physics explanations other than sterile neutrinos remain available to explain the Gallium Anomaly. 
% Recently, the PROSPECT collaboration published a proposal for an upgraded and improved detector, PROSPECT-II, which aims to probe the electron neutrino disappearance channel with high sensitivity at \dmTwo{}$\approx$1-10~eV$^2$, a partially unaddressed phase space region highly relevant to these new developments.  
% The PROSPECT-II detector features an evolutionary design which can be constructed and deployed within one year and have impactful results with as little as one calendar year of data.
% \end{abstract}

\maketitle
\snowmass{}

%\tableofcontents

\newpage

\section*{Executive Summary}

%\tlnote{TOM/PIETER UPDATE}

%(start with statement about role of reactors in HEP)

Nuclear reactors provide the highest intensity source of pure electron-type neutrinos available on earth. Reactor neutrino experiments
%Experiments leveraging nuclear reactors as extremely intense sources of man-made neutrinos
have played a central role in developing our current understanding of the three neutrino paradigm and in establishing the current era of precision neutrino physics. 
%Despite this, experimental anomalies continue to persist in measurements of both the flux and spectral shape of reactor neutrinos. 
%These anomalies may or may not be related to anomalies seen in other neutrino sectors and the global picture has been in significant tension for several years. 
%Above replace with Nathaniel's suggestion:
%Nonetheless, anomalous results persist across multiple sectors of neutrino physics, indicating the possibility of new physics. and specifically calling our understanding of reactor-emitted neutrino flux and spectrum into question.
Precision measurements of the flavor-pure antineutrino flux from reactors are one way to search for new physics by probing both the physics of neutrino oscillations and the production mechanism of rector antineutrinos. 
In the years to come reactor neutrino experiments will continue to play an important role in resolving the global neutrino picture. 
Unique features of reactor neutrinos mean that these experiments can continue to play a leading role in resolving the global neutrino picture. 

The PROSPECT experiment has substantially addressed the original `Reactor Antineutrino Anomaly' by  performing a high-resolution spectrum measurement from an enriched compact reactor core and a reactor model-independent sterile neutrino oscillation search based on the unique spectral distortions the existence of eV$^2$-scale sterile neutrinos would impart.
But as the field has evolved, the current short-baseline (SBL) landscape supports many complex phenomenological interpretations, establishing a need for complementary experimental approaches to resolve the situation.

While the global suite of SBL reactor experiments, including PROSPECT,  have probed much of the sterile neutrino parameter space, there remains a large region above 1 eV$^2$ that remains unaddressed. 
Recent results from BEST confirm the Gallium Anomaly, increasing its significance to nearly 5$\sigma$, with sterile neutrinos providing a possible explanation of this anomaly.
Separately, the \uB{} exclusion of electron-like signatures causing the \mB{} low-energy excess does not eliminate the possibility of sterile neutrinos as an explanation. 
In fact, \uB{} potentially indicates an oscillation-based deficit in the electron neutrino channel. 
Focusing specifically on the future use of reactors as a neutrino source for beyond-the-standard-model (BSM) physics and applications, higher-precision spectral measurements still have a role to play.

% The PROSPECT experiment was built to definitively address this situation by performing a high resolution spectrum measurement from an enriched compact reactor core and a sterile neutrino oscillation search based on the unique distortion of the energy spectrum as a function of baseline imparted by eV$^2$-scale sterile neutrinos. 
% While successful, neither PROSPECT nor the global suite of short-baseline (SBL) neutrino experiments have covered all of the relevant parameter space at high significance nor have spectral shape questions been resolved.
% With the release of new results from reactor experiments, BEST, and \uB{}, the search for sterile neutrinos has grown yet more complicated and interesting.

% While recent SBL reactor experiments have probed much of the sterile neutrino parameter space, there remains a large region above 1~eV$^2$ that is outstanding.
% The BEST experiment's recent results confirm the Gallium Anomaly and further increases its significance to nearly 5$\sigma$.
% No strong particle or nuclear physics explanations explain these results. 
% Separately, while \uB{} first results do not show an excess of electron-like signatures similar to that observed by \mB{}, their current sensitivity does not exclude all of sterile neutrino oscillation parameter space suggested by the \mB{} result.  
% In fact, in the electron neutrino channel, \uB{} potentially shows indications of an oscillation-based deficit.

These recent results have created a confusing landscape which requires new data to disentangle these seemingly contradictory measurements.
To directly probe \nuebar{} disappearance from high-\dmTwo{} sterile neutrinos, the PROSPECT collaboration proposes to  build an upgraded and improved detector, PROSPECT-II. 
It features an evolutionary detector design which can be constructed and deployed within one year and have impactful physics with as little as one calendar year of data.

\newpage

%\section{Compelling questions persist about the Short Baseline Oscillation Landscape after Recent Results}
\section{A Precision Reactor Oscillation and SPECTrum Experiment}

%\btfnote{(Moving this para to PRO1 section) }
The search for eV$^2$-scale sterile neutrinos is an active area of neutrino physics that is well motivated by theory and by experimental data.
%Previous experiments measuring inverse beta decay (IBD) \nuebar{} interactions from reactors found a global deficit of about 6\% compared to the state-of-the art reactor neutrino models including the Huber-Mueller~(HM model) model~\cite{Huber:2011wv,Mueller:2011nm}.
In the reactor neutrino sector in the early 2010s, roughly 6\% deficit was found in the experiments measuring inverse beta decay (IBD) \nuebar{} interactions from reactors compared to the then recently improved reactor neutrino flux models~\cite{Huber:2011wv,Mueller:2011nm}.
This discrepancy, referred to as the Reactor Antineutrino Anomaly (RAA), hinted at the possible existence of a sterile neutrino flavor~\cite{Mention:2011rk}.
Additionally, radiochemical solar neutrino experiments based on gallium found a $\sim$3$\sigma$ deficit of detected $\nu_e$ interactions from nearby intense radioactive sources, referred to as the Gallium Anomaly (GA)~\cite{Giunti:2010zu}. 
These two anomalies have prompted an intense global experimental campaign using MeV-scale neutrino sources to test for the existence of sterile neutrinos.% and understand the underlying physics.  

Built in part to provide a definitive search for sterile neutrino oscillations at very short baselines, the PROSPECT experiment was supported by the Intermediate Neutrino Research Program~\cite{Adams:2015ogl} that followed from the 2014 P5 report. 
PROSPECT provided strong constraints on sterile neutrinos over significant portions of the phase space suggested by the RAA as well as a high-resolution measurement of neutrinos from an HEU core. 
While successful in supporting many of the collaboration's science goals, the PROSPECT detector suffered from technical problems which cut short its useful life.
As described below, the short baseline oscillation landscape continues to evolve, motivating the PROSPECT collaboration to preparing for an evolutionary detector upgrade (PROSPECT-II) that builds from the success of the experiment so far and leverages that existing investment. 
The PROSPECT-II upgrade, which is described in detail in Ref.~\cite{Andriamirado:2021qjc}, resolves technical issues that abbreviated the first run, introduces design features that improve robustness and time-stability, and extends both the depth and the scope of the experiment’s physics reach.

\section{Recent results further complicate the Short Baseline Oscillation Landscape}

% Compelling Questions persist about the Short Baseline Oscillation Landscape after Recent Results

% \begin{itemize}
% \item VSBL measurements (PROSPECT, STEREO, NEOS, DANNS, Nu4) 
% \item \ptsnote{I also suggest that we briefly describe Daya Bay and RENO evolution results here. something along these lines 
% `The results from Daya Bay~\cite{} and RENO~\cite{} evolution analyses have measured \uFive{} flux lower than the Huber-Mueller model~\cite{giunti_diagnose}.'} 
% \item \uB{} and the LEE null-observations
% \item BEST strengthening the Gallium Anomaly
% \end{itemize}

% Motivation: Historical RAA & GA

%Recent Reactr expt development
A number of experiments including PROSPECT, STEREO, NEOS, DANSS, and Neutrino-4 have probed $\nuebar$ oscillations at very short baselines from reactors~\cite{PROSPECT:2020sxr,STEREO:2019ztb,NEOS:2016wee,DANSS:2018fnn,NEUTRINO-4:2018huq,Serebrov:2020kmd}.
Each experiment uses model-independent spectral ratio measurements which directly search for energy and baseline dependent spectral distortions that are unique to sterile neutrino oscillations. 
With the exception of Neutrino-4, the experiments' results have been found to be statistically consistent with the three neutrino model.  
Neutrino-4 reports evidence for sterile oscillation with 2.9$\sigma$ significance\footnote{For consistency, this note uses the published Neutrino-4 results from Ref.~\cite{Serebrov:2020kmd}.}, but is in direct tension with the other reactor experiments and the analysis has drawn criticism~\cite{Giunti:2021iti,PROSPECT:2020raz,Coloma:2020ajw}.
Overall, these direct oscillometry experiments have excluded large portions of low-$\Delta$m$^2$ preferred regions for RAA and GA.

%BRL: I think it might make sense to have a paragraph break here.  Then the text below is proving storyline development in a different direction -- above) we used clear osc seraches and didn't see anything; below) we also have some flux model input developments that tell us maybe the RAA is from something other than steriles (berryman, giunti papers) BUT flux developments alone are unlikely to have the statistical power to authoritatively exclude the possibility of a sterile-induced deficit in reactor fluxes.  Also should mention DYB evolution and summation calcs if you're gonna have this discussion here.  KI alone I think isn't too convincing.
%BRL: Alternatively, I think you could ditch ALL of the text below, and move it to the 'pheno' section, but could at the end of the previous paragraph mention that theta13 neutrino data has suggested issues with flux and spectrum modelling, providing a promising 'non-BSM' avenue for explaining the RAA.
%Recent flux model development
As the RAA is based on an observed deficit between the predicted and measured \nuebar{} fluxes at multiple reactor sites, it depends on the accuracy of reactor flux predictions.
These predictions are based on neutron-induced fission beta spectra collected by Schreckenbach et al. at the Institut Laue-Langevin (ILL) and converted into neutrino spectra by Huber~\cite{Huber:2011wv} and Mueller~\cite{Mueller:2011nm} (referred to as the HM model).
A recent Kurchatov Institute (KI) measurement of the ratio of the cumulative $\beta$-decay spectrum between \uFive{} and \pNine{} is lower than the ILL/HM value by 5.4\%~\cite{Kopeikin:2021rnb}.
It was suggested by Kopeikin et al., that this discrepancy is likely due to an overestimation of the absolute normalization of \uFive{} at ILL.
Under this assumption, the re-evaluated flux (KI model) based on the modified normalization produces IBD yields that agree with reactor flux and evolution measurements within 1$\sigma$~\cite{Kopeikin:2021ugh}, reducing the significance of the original motivation for the RAA.
Though is appears likely that a normalization error contributed to the original RAA, the KI model does not preclude sterile neutrinos from existing in this parameter space as shown in Fig.~\ref{fig:exclusions}.

New results from SBL reactor experiments, BEST, and \uB{} have brought new information and interest in a potential eV-scale sterile neutrino.
%BEST
In contrast to the RAA, the GA requires no reactor flux prediction or knowledge.
The initial gallium experiments SAGE and GALLEX were not purpose-built to probe for sterile neutrinos. 
To directly probe the GA, the Baksan Experiment on Sterile Transitions (BEST) measured the rate of neutrino interactions in a layered gallium detector, with a high intensity $\nu_e$ source in the center~\cite{Barinov:2021asz}.
To search for oscillations, the rate of production of $^{71}$Ge is measured in inner and outer volumes and compared to expected results. %, as shown in Fig~\ref{fig:Exp_1}.
The BEST results show a $\sim$20\% deficit in both volumes, strengthening the significance of the GA, but not providing any  indication as to whether the deficit is oscillatory in nature.  

%MicroBoone
Neutrino experiments using accelerators have provided intriguing short-baseline anomalies, and remain a highly active avenue for probing sterile oscillations.  
In the 1990s and 2000s, accelerator neutrino measurements by LSND and \mB{} found an excess of $\nu_e$-like and \nuebar{}-like events from predominantly $\nu_\mu$ sources, with the \mB{} excess eventually established with 4.8$\sigma$ statistical significance~\cite{LSND:2001aii,MiniBooNE:2020pnu}.  
Potential explanations of these results have involved sterile neutrinos, other BSM physical phenomena, or some combination of the two.
The corresponding sterile oscillation for this anomaly is in a similar region of the \dmTwo{} parameter space as the RAA and GA, increasing interest in a sterile neutrino of this scale.
%BRL: mention that dm2 ranges for explaining this anomaly are similar to those used to explain RAA and GA. BTF: done
Recent results from \uB{} using a beam-line and baseline very similar to \mB{} show no such excess~\cite{MicroBooNE:2021rmx}, though their initial sensitivity does not cover the entirety of the \mB{} suggested region.
Interestingly, \uB{} observes a modest deficit in measured $\nu_e$~\cite{MicroBooNE:2021zai}, which some interpret as a hint of BSM physics~\cite{Denton:2021czb}. % as seen in Fig~\ref{fig:Exp_1} 

% \begin{figure}
%     \centering
%     \includegraphics[width=0.45\textwidth]{figures/MuB_Wire_Cell.png}~
%     \includegraphics[width=0.45\textwidth]{figures/BEST_Rate.png}
%     \caption{
%     \btfnote{Currently using figures representing MicroBoone and BEST results. Could instead use exclusion curves, but overlaps with Fig \ref{fig:exclusions}. Could use a plot summarizing VSBL reactors instead of one of the other 2.}
%     Left: ~\cite{microboonecollaboration2021search}. 
%     Right: ~\cite{barinov2021results}
%     }
%     \label{fig:Exp_1}
% \end{figure}

%==============================================================================

\section{Complementary Experimental Approaches to resolve potential phenomenological explanations}

\begin{figure}[h]
    \centering
    \includegraphics[trim = 0cm 0.0cm 0.5cm 0.0cm, clip=true, height=2.7in]{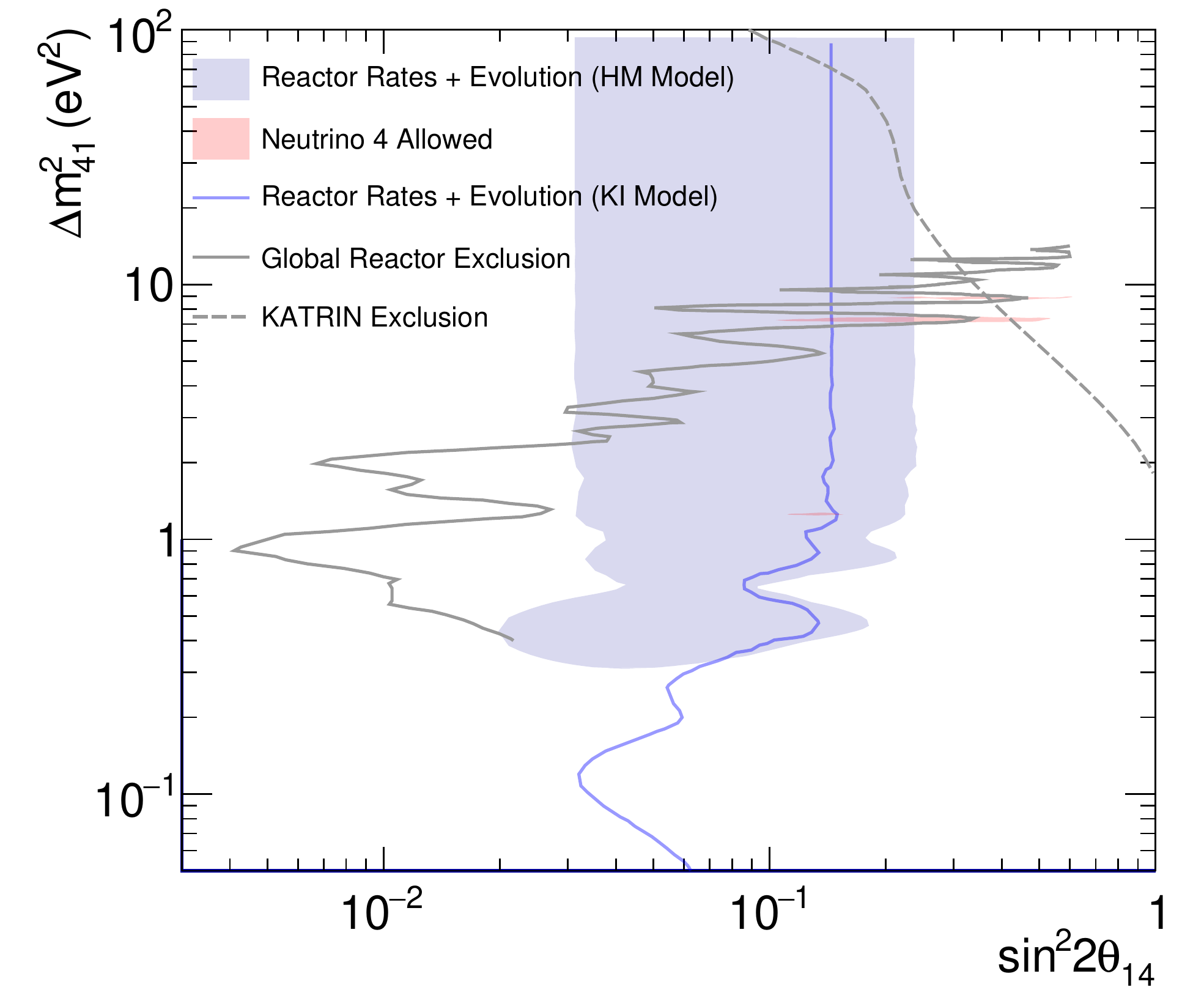}
    \includegraphics[trim = 0cm 0.0cm 0.5cm 0.0cm, clip=true, height=2.7in]{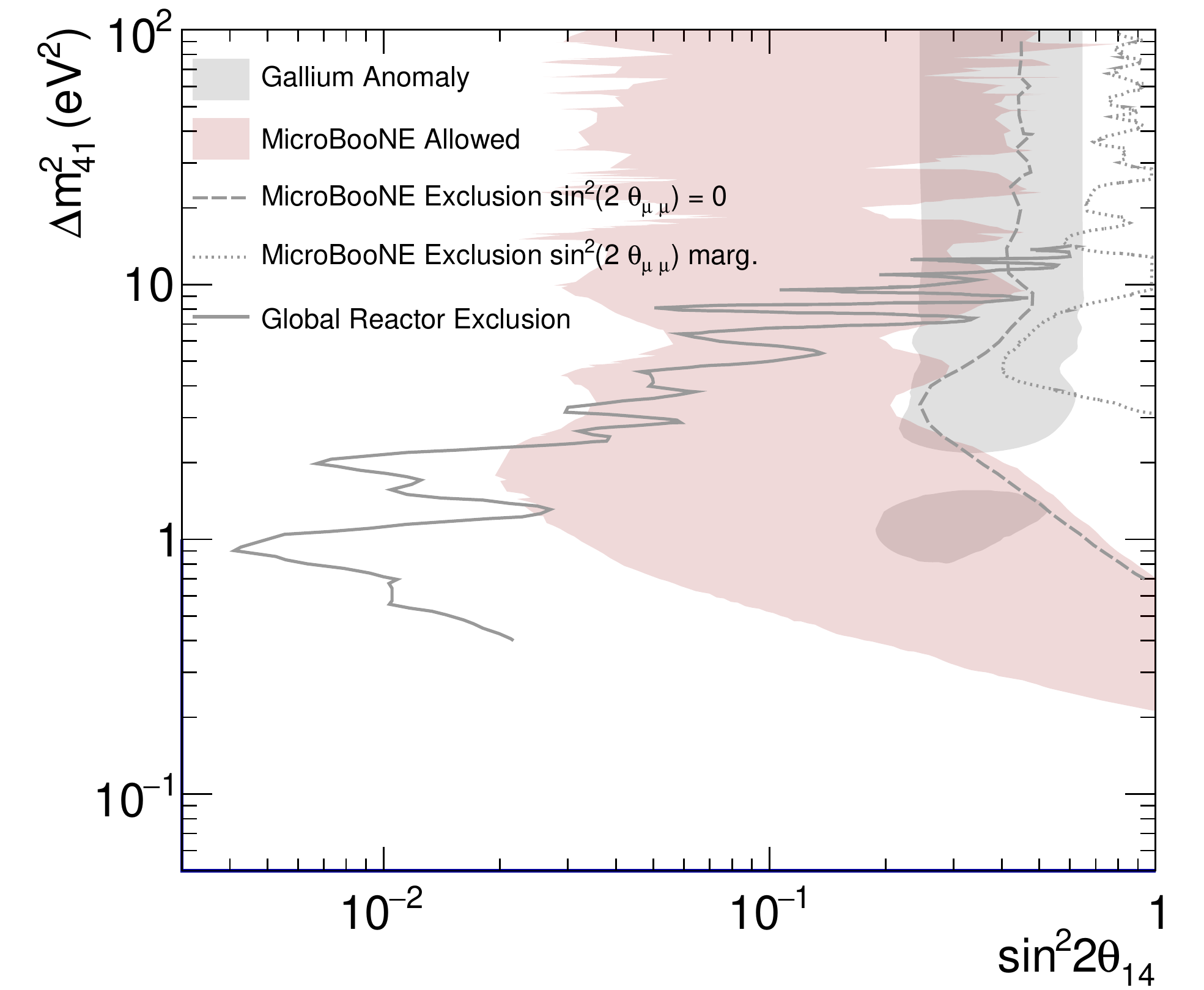}
    \caption{\textbf{Left:} Comparison of the suggested parameter space from RAA (HM model)~\cite{Giunti:2021kab} and Neutrino-4~\cite{Serebrov:2020kmd} to the allowed regions from the RAA (KI model)~\cite{Giunti:2021kab} and excluded parameter regions from global fits of spectral-ratio reactor measurements~\cite{Berryman:2021yan} and KATRIN experiment~\cite{Aker:2022ldk}. \textbf{Right:} Comparison of the suggested parameter space from the gallium anomaly~\cite{Barinov:2021mjj} and two $\nu_{e}$-disappearance analyses using \uB{} data, one hinting~\cite{Denton:2021czb} at oscillations and the other~\cite{Arguelles:2021meu} excluding a small portion of the parameter space, to the excluded parameter regions from global fits of spectral-ratio reactor measurements~\cite{Berryman:2021yan}. Both cases show regions of interesting parameter space with $\Delta m^2 > 5 $ eV$^{2}$ yet to be explored.}
    \label{fig:exclusions}
\end{figure}
%BRL: Can we split this one into 'MeV-scale-and-below sector' (i.e. RxSpectrumRatios, RAA, GA, Katrin on one) and 'accelerator sector (RxSpectrumRatio, uB1, and uB2 on the other)?  It just seems quite busy right now, and in the paragraphs below, the narrative seems somewhat split out in this manner anyways.
%PTS: Done

It is worth considering the aforementioned experimental results in a broader phenomenological context to inform future experimental efforts. 
There is an increasing amount of evidence suggesting that the source of RAA is, at least in part, due to the mismodeling of the reactor \nuebar{} spectra--primarily driven by \uFive{}.
This interpretation is supported by an improved agreement between the measured isotopic IBD yields and the new updated summation model~(the Estienne-Fallot or EF model) based on the revised nuclear databases~\cite{Estienne:2019ujo} along with the Daya Bay~\cite{DayaBay:2017jkb,DayaBay:2019yxq} and RENO~\cite{RENO:2018pwo} fuel evolution results and re-evaluated KI-based conversion model.
Combined fits of the reactor antineutrino yields and the Daya Bay and RENO evolution data-sets suggest a persistent RAA at $\sim 3\sigma$ when compared to the ILL/HM model while the anomaly reduces to $\sim 1\sigma$ when compared to the KI and EF models~\cite{Giunti:2021kab}.

When considered in the context of a 3+1 sterile neutrino hypothesis, the EF and KI models have no statistically significant preference for eV-scale oscillations.
%the HM model continues to have $\sim 2\sigma$ bounded regions, as shown in the Fig~\ref{fig:exclusions}, whereas 
% All the reference models have upper bounds of $ 0.14<\sin^{2}{\theta_{ee}}<0.25$ for \dmTwo$_{14} > 1 \mathrm{eV}^2$, pointing to the parameter space still to be addressed by the upcoming experiments.
Though the sterile neutrino explanation of the RAA is diminished with the updated models, the combined reactor rate and evolution data do not preclude the presence of sterile neutrinos in this region, as shown on the left panel of Fig.~\ref{fig:exclusions}. 
Viable hybrid models exist that could accommodate incorrect reactor neutrino flux predictions while also allowing oscillations to sterile neutrinos~\cite{Giunti:2019qlt}.
Rate and flux-evolution measurements alone are not sufficient to unambiguously resolve the reactor anomaly, due in part to reactor power uncertainties and the complex uncertainties in predicting neutrino spectra from fission.
Relative spectral measurements, such as those deployed in SBL reactor experiments, are needed for a definitive resolution.
%BRL: should also mention 'hybrid models,' which also give a lot of wiggle room as to allowed sterile osc parameters.
%DONE\

Over the past 5 years SBL reactor experiments performing oscillation searches using relative spectral measurements have collected considerable amount of valuable data.
A combined analysis~\cite{Berryman:2021yan} using data from these SBL experiments--including PROSPECT~\cite{PROSPECT:2018dtt}, STEREO~\cite{STEREO:2019ztb}, NEOS~\cite{NEOS:2016wee}, DANSS~\cite{DANSS:2018fnn}, and Neutrino-4~\cite{NEUTRINO-4:2018huq}--shows no strong evidence of sterile neutrino oscillations at the eV-scales.
The use of relative oscillation searches for this combined fit makes it robust against reactor modeling uncertainties. % and demonstrates the power of relative oscillation searches offered by the SBL reactor experiments.
As shown in Fig.~\ref{fig:exclusions}, while the combination of these experiments excludes major portions of sterile neutrinos parameter space, a sizeable fraction of the RAA still persists.
Moreover, these results are compatible with the gallium results under a 3+1 sterile neutrino model at similar oscillation frequencies. % at \dmTwo{} $>$ 5 eV$^2$.
Hence, more data covering \dmTwo{} $>$ 1 eV$^2$ are needed to fully explore this parameter space.
% \blnote{should we also state that they also are consistent with a sterile-induced RAA above roughly 3 $eV^2$}
% \ptsnote{Hmm, I am conflicted. This may give way too much credence to N4 results and PROSPECT-II can't really address much of that region in the lower s22t anyway.}
% \ptsnote{Should we talk about solar experiments ?, they disagree with these results.}
%BRL: I would move the points beneath to the final paragraph of this section, where you're ballooning out to discussion of 'non-vanilla BSM.'  Right here, it just kinda seems dropped in.  You can just let this paragraph be about 'if RAA and GA are real and from osc, our unambiguous spectral ratio experiments have yet to test all the relevant space AND this is particularly true at higher dm2, a parameter space range only accessible to very short-baseline compact core reactor experiments AND that neutrino mass experiments, which push from the high dm2 side, won't resolve this space anytime soon.  
%DONE

The first data release of the \uB{} experiment has not shown indications of $\nu_e$ appearance from $\nu_{\mu} \rightarrow \nu_{e}$ oscillations. 
Nevertheless, the presence of intrinsic $\nu_{e}$ in the beam allows for a $\nu_{e}$ disappearance search with this dataset.
One such preliminary analysis~\cite{Denton:2021czb} performed using the \uB{}'s data release hints at a 2.2$\sigma$ evidence of sterile neutrinos in the similar parameter space as the RAA and the GA with the best-fit point at ($\mathrm{sin}^2(2\theta_{14}) = 0.30$ and $\Delta m^2_{41} = 1.42 $eV$^2$).
A more rigorous fully-consistent 3+1 neutrino oscillation based approach~\cite{Arguelles:2021meu} using the same \uB{} dataset but including the official \uB{} covariance matrix\footnote{Note that neither of these analyses are performed by the \uB{} collaboration and the outcomes from the official analysis may vary slightly.} that accounts for correlated systematic uncertainties sees no hints of oscillations. 
The results from both these are shown in the right panel of Fig.~\ref{fig:exclusions}.
%BRL: Also mention that they considered both nue appearance and nue dis, which are degenerate effects that complicate sterile searches for these impure sources.  
This analysis excludes portions of parameter space suggested by the \mB{}, reactor, and gallium anomalies.  
The final \uB{} dataset, planned to be $\sim$2x the size of the current analyzed dataset, is expected to improve the experiment's coverage but significant portions of the $\sin^2\theta_{e e}$ parameter space will remain unexplored.
The presence of $\nu_{e}$ in the ~\uB s $\nu_{\mu}$ beam produces degenerate effects between $\nu_{e}$ disappearance and $\nu{e}$ appearance complicating the interpretations for sterile neutrino oscillation searches.
These results highlight the importance of a flavor-pure neutrino source and the need for complementary sterile neutrino searches that can fully address the parameter space suggested by all anomalies shown in Fig.~\ref{fig:exclusions}.
%BRL: A KEY KEY KEY missing point here is that nue appearance and nue dis are degenerate effects that complicate sterile searches from impure sources.  
%DONE

Looking at the broader picture, \uB{} results so far don't resolve the decades-long \mB{}~\cite{MicroBooNE:2021zai} and LSND~\cite{LSND:1997vun,LSND:2001aii} anomalous results. 
Moreover, the reconciliation of LSND, \mB{}, and \uB{} results demand invocation of a combination of multiple non-vanilla BSM models.
The picture gets even more complicated when datasets from reactor and gallium experiments are included.
A key point to note is that while the gallium experiments were so far only able to probe the deficit and can't disambiguate between effects arising from oscillations or an unknown production effect, relative reactor searches have the powerful capability to directly search for the propagation effect induced by neutrino oscillations.
Ultimately, the consolidation of these paradoxical results necessitates the need for multiple complementary probes to  disentangle multiple competing BSM effects.
%BRL: I think you need a paragraph here about results from last 10 years highlighting that non-vanilla BSM kinda seems necessary to reconcile various SBL results.  In here you can: a) mention the long-standing dis-app tensions; b) state that current uB and mB results can be more easily reconciled if multiple effects cause the mB anomaly; c) mention that GA could be production BSM, while reactors probe propagation BSM; d) end by highlighting that multiple channels, energies and flavor purities can enabling disentangling of multiple competing BSM effects.  

%BRL: We are also missing our canonical 'CPV disambiguation paragraph' which remains a key plank of our physics case.  Should be inluded?  or no?  Could add this as a single sentence in your section conclusion below?

Despite significant experimental, theoretical, and phenomenological progress in the reactor, gallium, and long baseline sectors, a consistent description of the neutrino picture hasn't emerged yet.
% Recent results from the \uB{} and BEST experiments 
%in conjunction with the reactor neutrino datasets leaves the anomalies unresolved and necessitates new experimental data.
% As shown in the shaded regions in Fig.~\ref{fig:exclusions}, various anomalous results suggest active-sterile mixing with \dmTwo{}$ > 1 \mathrm{eV}^2$ and demands robust oscillation measurements with access to high frequency oscillations.
The combined picture of all the anomalous results cannot be fully explained using a 3+1 sterile neutrino picture highlighting the need for multiple complementary efforts to comprehensively probe the anomalies.

%BRL: This last summary paragraph here needs to be sharpened -- the main points should be more clear and organized: a) Despite a major advancement in available BSL data in last 10 years, a harmonious picture, whether SM or BSM-including, has not coalesced.  b) in part, this is because we're missing data in key areas: despite developments in reactor models and huge expansions in available reactor spectral ratio data, a clear region of unaddressed oscillation parameter space exists in the electron disappearance sector from roughly 3-20 eV^2.  c) in part, it's also because we don't have diverse enough datasets to probe the interplay of multiple possible BSM effects.  d) A global SBL program for the next 10 years should seek to remedy both of these primary issues. 

%==============================================================================

%\section{PROSPECT-II Oscillation Physics Opportunities}
\section{PROSPECT-II: unique inputs for the resolution of short baseline anomalies}

% \begin{itemize}
%     \item Complementary to \uB{}/SBND/accelerator searches
%     \item Physics goals: high $\Delta$m$^2$, kill or confirm the hints from Nu4 and BEST
%     \item Disagreement b/w reactor and Ga anomaly may also point to production vs oscillation effect. PROSPECT will probe the oscillation effect
%     \item reference PII physics paper for experimental details
% \end{itemize}

\begin{figure}[h]
\centering
\includegraphics[trim = 0cm 0.0cm 0.5cm 0.0cm, clip=true, height=2.7in]{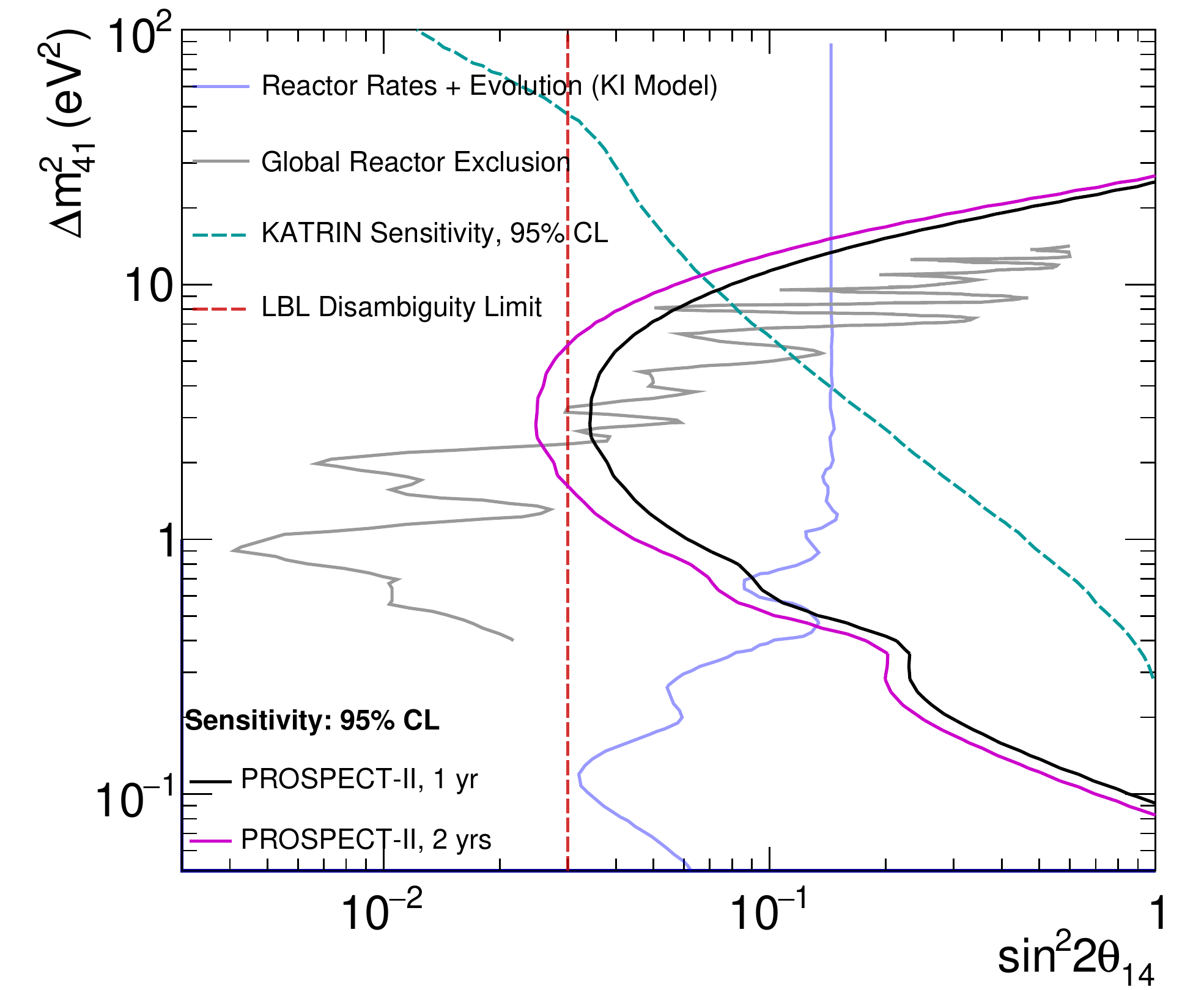}
    \includegraphics[trim = 0cm 0.0cm 0.5cm 0.0cm, clip=true, height=2.7in]{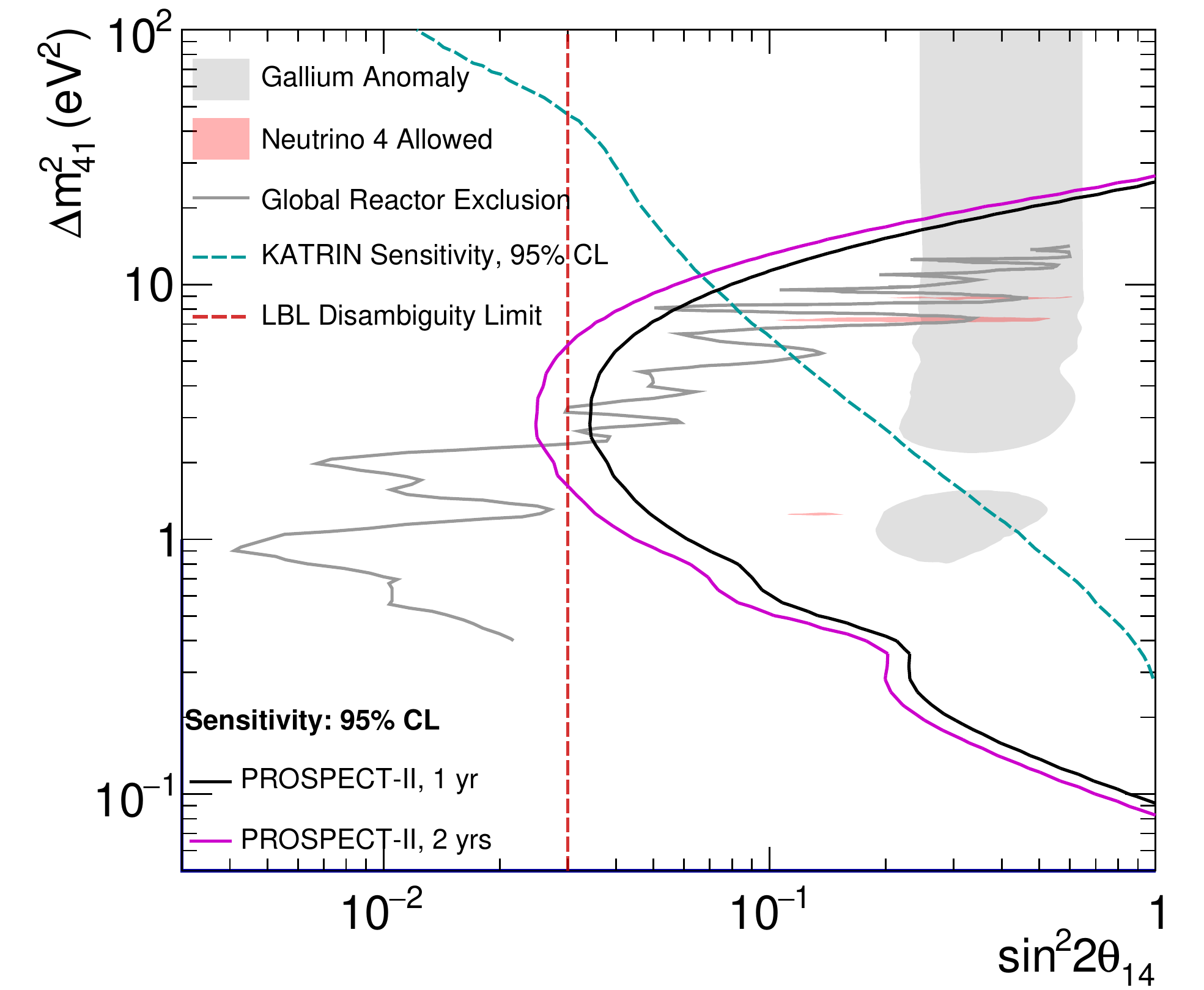}
\caption{Sensitivity contours from one year~(black, solid) and two years~(pink, solid) of PROSPECT-II~\cite{Andriamirado:2021qjc} data-taking compared to: \textbf{Left}: Already excluded parameter space from the relative reactor spectral experiments~(gray, dashed), and the allowed region~(blue, solid) from RAA (KI model), \textbf{Right:} Suggested parameter space from GA and Neutrino-4 experiment~(pink) and the CP violation disambiguity limit (red, dashed). PROSPECT-II can significantly increase the global sensitivity in the 1-10~eV$^2$ range. Additionally, PROSPECT-II in conjunction with the projected sensitivity~(dashed, teal) from KATRIN~\cite{Aker:2022ldk} will be able to exclude all of the GA suggested parameter space and clear up the CP violation disambiguity.}
%BRL: Again, I'd separate out the accelerator and MeV-scale-and-below sectors?

\label{fig:sensitivity}
\end{figure}

In light of these recent developments, the physics opportunities for PROSPECT-II become even more tantalizing. 
In 2021, the PROSPECT collaboration published a detailed summary of the physics opportunities with an upgraded detector which can be rapidly deployed~\cite{Andriamirado:2021qjc}.
As detailed above, both \uB{} and the gallium experiments point to preferred parameter space in the few-eV$^2$ region, with oscillation amplitudes just beyond what has been probed by the current generation of SBL reactor experiments.

Reactor experiments are highly complementary to accelerator- and source-based measurements and feature a flavor-pure, high-intensity source of \nuebar{}.
While there have been questions about the uncertainties in absolute flux predictions, segmented detectors at short baselines are able to directly search for energy-dependent oscillation-induced spectral distortions that are the `smoking-gun' of sterile neutrinos.
This model-independent technique is crucial to positively identify neutrino oscillations as opposed to an ambiguous flux-deficit that could be caused by a mismatch between data and theoretical predictions. 
The energies ($\sim$few MeV) and baseline (7-9~m) available to PROSPECT-II at HFIR are uniquely suited to searching for oscillations in the 1--10~eV$^2$ region.
The projected PROSPECT-II sensitivity will surpass the current global analysis' precision at all $\Delta$m$^2$ above  $\sim$2~eV$^2$ with as much as 2 to 4 times improvement for mass-splittings in the 5--10~eV$^2$ region.

The Neutrino-4 collaboration reports a $\sim$3$\sigma$  oscillation-like signal with a best-fit point of $\sim$7.3~eV$^2$ and an amplitude of $\mathrm{sin}^22\theta = 0.36$.
The allowed region is shown in Fig.~\ref{fig:sensitivity}.
This best-fit point is in tension with results from PROSPECT and STEREO.
By probing this broad-region of parameter space, PROSPECT-II can play a valuable role in the resolution of the aforementioned confusing experimental and theoretical landscape.

While testing the presence of an additional sterile state is an important BSM study in its own right, it is also crucial for the future studies of Standard Model neutrino parameters. 
Results from upcoming long baseline~(LBL) experiments designed to measure CP-violation remain ambiguous~\cite{deGouvea:2014aoa,Gandhi:2015xza,deGouvea:2016pom,Klop:2014ima} if sterile neutrinos are not fully excluded for mixing angles $\sin^2 2\theta \gtrapprox 0.03$~\cite{,Dutta:2016glq}.
Thus a combination of PROSPECT-II, tritium beta endpoint measurements, and medium baseline neutrino experiments together will play a complementary role in the interpretation of the future LBL results.

%BRL: I think this should be split into two paragraphs.  The first is about 'clarifying the MeV-and-below sector' and the second is about 'clarifying the accelerator sector.'  For MeV and below: a) we can address N4; b) combined with KATRIN, we address entire canonical RAA and GA regions at ALL relevant dm2.  c) in concert with GA, P2 can generate experimental impetus for  development of new theory (BSM or non-BSM) explanations for the GA.  For accelerator: a) we can pin down nue dis so that SBL nue appearance results can perform more unambiguous nue app 3+1 osc fits, as opposed to resorting to combined analyses with beams of differing nue purity; and b) we can pin down 'vanilla 3+1' behavior so that these experiments have more unambiguous sensitivity to other physics in this hidden sector, such as decay, etc. etc. etc.  
%BRL: Finally, we need to mention the PROSPECT2 is the ONLY ONLY ONLY experiment on the horizon capable of meeting these needs for the neutrino community!  It's us or nothing!!!!

%\section{PROSPECT-II ``other'' Scientific Opportunities}

\section{PROSPECT-II: Benchmark measurements for precise understanding of reactor \nuebar{} emission}

There are additional scientific goals that the PROSPECT-II upgrade can achieve. These primarily relate to greatly improving our understating of reactors as an antineutrino source which would benefit neutrino physics, BSM studies, and safeguards applications using neutrinos~\tlnote{Cite reactor whitepaper}. 

Comparisons of experimental and predicted \nuebar{} energy spectra measured at LEU-fueled reactors show sizable disagreements, most prominently in the 4-6~MeV energy range.  
PROSPECT-II will help to address this situation by further improving the precision of the world-leading PROSPECT measurement of the $^{235}$U \nuebar{} energy spectrum.   
As described in Ref.~\cite{Andriamirado:2021qjc}, PROSPECT-II will produce a spectrum  measurement that approaches or exceeds the precision of current prediction approaches, providing a stringent test of the underlying models and nuclear data.   
Furthermore, a joint analysis of spectrum measurements from Daya Bay and PROSPECT-II would produce purely data-driven reactor $\overline{\nu}_e$ spectrum models for future particle physics measurements and potential applications. 
Benchmark spectra have been identified as a high-priority ``nuclear data'' need during a recent community workshop~\cite{bib:WoNDRAM}.

PROSPECT-II will also perform a precise measurement of the \nuebar{} flux produced in \uFive{} fission.  
%Similar to the situation with the reactor \nuebar{} spectrum, measurements and flux prediction are discrepant. 
By performing a modern \uFive{}~\nuebar{} flux measurement, PROSPECT-II can increase the reliability of the global flux picture,
similarly benefiting the particle physics and nuclear science communities. A flux precision of 2.5\% is anticipated, with the dominant systematic being knowledge of the HFIR power ($\sim$2$\%$). When combined with flux measurements at LEU fueled reactors that have  a more complex fuel mix, the pure \uFive{}~\nuebar{} flux measurement performed by PROSPECT-II  would improve the precision of IBD yields from all major fissioning isotopes~\cite{Andriamirado:2021qjc}.

%\section{The PROSPECT-II Experimental Program}

\section{PROSPECT-II: An evolutionary detector upgrade with physics results within 2 years}
 
%  (Tom and Nathaniel)
 
%  \begin{itemize}
%      \item P-I Detector accomplishments
%      \item P-II Detector evolutionary upgrade
%      \item P-II timeline 
%  \end{itemize}

% \ptsnote{Explicitly mention that it came out the last P5 and intermediate program.} 

% PROSPECT-II is an evolutionary experiment that builds upon the successes of the PROSPECT-I program while revising the detector design to increase longevity and stability.
% The active detector is comparable in size ($\sim$4000~kg) and expected performance and is designed to be deployed in the same experimental hall as PROSPECT utilizing existing infrastructure and shielding package.
% A detailed summary of the PROSPECT-II detector and proposed operation can be found in Ref.~\cite{Andriamirado:2021qjc}.

%The PROSPECT experiment was supported by the Intermediate Neutrino Research Program~\cite{Adams:2015ogl} that followed from the 2014 P5 report. 
%The PROSPECT collaboration is preparing for an evolutionary detector upgrade (PROSPECT-II) that builds from the success of the experiment so far and leverages that existing investment. 
%The PROSPECT-II upgrade, which is described in detail in Ref.~\cite{Andriamirado:2021qjc}, resolves technical issues that abbreviated the first run, introduces design features that improve robustness and time-stability, and extends both the depth and the scope of the experiment’s physics reach.

The original PROSPECT detector initially met all design requirements, as laid out in Ref.~\cite{PROSPECT:2015iqr}.
Unprecedented background rejection, provided by detector segmentation and particle identification via Pulse Shape Discrimination using a $^6$Li-doped liquid scintillator~\cite{PROSPECT:2018dnc,PROSPECT:2019enz}, allowed precision reactor antineutrino measurements to be conducted near the earth's surface with very little overburden. 
Excellent energy resolution, precision energy calibration and reconstruction, and event position reconstruction~\cite{PROSPECT:2018hzo} in a compact detector enabled model-independent short baseline oscillation searches and modern antineutrino energy spectrum measurements from $^{235}$U fission~\cite{PROSPECT:2018dtt, PROSPECT:2018snc, PROSPECT:2020sxr}.

As in the original PROSPECT design, the PROSPECT-II detector will contain a segmented $^6$Li-doped liquid scintillator volume optimized for inverse beta decay detection with minimal cosmic-ray shielding. 
The PROSPECT-II detector design addresses technical issues encountered during the initial data taking period that caused a fraction of the  detector PMTs to become inoperable.
%by introducing evolutionary changes that require modification to only a minority of subsystems, and that maintain the performance required to achieve the physics goals of the experiment.  
The principal design change moves the PMTs outside the liquid scintillator volume, thereby eliminating the possibility of liquid scintillator affecting voltage divider operation. 
Additionally, this change reduces the range of materials in contact with the liquid scintillator, providing an improved environment for long-term stability and operation. 
Completing the upgrade involves rebuilding the inner scintillator containment vessel, the production of new liquid scintillator, and a revamped calibration deployment scheme.  
Components outside this inner region, including an outer liquid containment vessel, an extensive shielding package and data acquisition electronics are largely unchanged.  
These evolutionary changes require modification to a minority of subsystems and are expected to maintain the demonstrated performance achieved during initial PROSPECT operation. 

Based on the demonstrated construction timeline of PROSPECT, the PROSPECT-II detector can be built and deployed within one calendar year of project start. 
This ability to leverage existing components and expertise makes it possible for PROSPECT-II to rapidly begin collecting the largest ever data set from an isotopically pure source of \uFive fissions at the High Flux Isotope Reactor.
Impactful physics results can then be produced with as little as one calendar year of data, with full sensitivity being reached after 14 reactor cycles (Fig.~\ref{fig:sensitivity}). With a timely start, this can be comfortably achieved prior to a long reactor outage planned for 2028~\cite{HFIR-schedule}.

\section{Summary}
% \begin{itemize}
%     \item Recent results from \uB{} and BEST enhance the need for a VSBL reactor program specifically at higher dm2
%     \item PROSPECT-II will be the experiment to cover 1-10 eV2
%     \item KATRIN and P8 will cover $\Delta m^2 > 10eV^2$ and a combination with PROSPECT will cover several generations in dm2
%     \item PROSPECT-II is ideally suited to resolve this if we move fast
%     \item Time is of the essence, HFIR shutting down soon
% \end{itemize}

%In this short note, we have detailed recent experimental and phenomenological results advancing our understanding of the eV-scale sterile neutrino landscape.
Short-baseline reactor experiments have been very successful in probing low-mass ($<$1eV$^2$) sterile neutrinos, though sensitivity to the high-\dmTwo{} region remains limited.  
\uB{} and BEST have generated renewed excitement about the possibility of high-\dmTwo{} sterile neutrinos.
Efforts to interpret these results have demonstrated the need for new and enhanced data that can probe this region.
The KATRIN experiment is beginning to probe the $>10$~eV$^2$ region and the $\theta_{13}$ reactor experiments have effectively covered the low-$\Delta m^2$ region, leaving an opportunity for short-baseline reactor-based experiments to probe for 1--10~eV$^2$ mass-splittings.
We highlight the unique contributions that the recently proposed PROSPECT-II physics program can make to this exciting landscape.
By rapidly deploying a robust detector, it is possible to explore this region for new physics in a two-year timeline. 

\section*{Acknowledgements}

The PROSPECT experiment is supported by the following sources: US Department of Energy (DOE) Office of Science, Office of High Energy Physics under Award No. DE-SC0016357 and DE-SC0017660 to Yale University, under Award No. DE-SC0017815 to Drexel University, under Award No. DE-SC0010504 to University of Hawaii, under Award No. DE-SC0008347 to Illinois Institute of Technology, under Award No. DE-SC0016060 to Temple University, under Contract No. DE-SC0012704 to Brookhaven National Laboratory, and under Work Proposal Number SCW1504 to Lawrence Livermore National Laboratory. This work was performed under the auspices of the U.S. Department of Energy by Lawrence Livermore National Laboratory under Contract DE-AC52-07NA27344 and by Oak Ridge National Laboratory under Contract DE-AC05-00OR22725. Additional funding for the experiment was provided by the Heising-Simons Foundation under Award No. \#2016-117 to Yale University. J.G. is supported through the NSF Graduate Research Fellowship Program and A.C. performed work under appointment to the Nuclear Nonproliferation International Safeguards Fellowship Program sponsored by the National Nuclear Security Administration’s Office of International Nuclear Safeguards (NA-241). This work was also supported by the Canada First Research Excellence Fund (CFREF), and the Natural Sciences and Engineering Research Council of Canada (NSERC) Discovery program under grant \#RGPIN-2018-04989, and Province of Ontario. We further acknowledge support from Yale University, the Illinois Institute of Technology, Temple University, Brookhaven National Laboratory, the Lawrence Livermore National Laboratory LDRD program, the National Institute of Standards and Technology, and Oak Ridge National Laboratory. We gratefully acknowledge the support and hospitality of the High Flux Isotope Reactor and Oak Ridge National Laboratory, managed by UT-Battelle for the U.S. Department of Energy.

\bibliography{refs}% Produces the bibliography via BibTeX.

\end{document}